\documentclass[useAMS,usenatbib]{mn2e}
\usepackage{graphicx,natbib,color,multirow,amsmath}
\definecolor{titlecol}{rgb}{0,0,1}
\definecolor{titlecol2}{rgb}{0,0.65,0}

\def\changed    {}

\def\oiii		{$\mathrm{\left[ O \textsc{iii}\right] }$}

\def\nii		{$\mathrm{\left[ N \textsc{ii}\right] }$}

\def\sii		{$\mathrm{\left[ S \textsc{ii}\right] }$}

\def\galfit     {{\tt GALFIT}}

\def\mmsun	{\rm{M}_{\odot}}

\def\lesssim{\mathrel{\hbox{\rlap{\hbox{\lower3pt\hbox{$\sim$}}}\hbox{\raise2pt\hbox{$<$}}}}}
\def\gtrsim{\mathrel{\hbox{\rlap{\hbox{\lower3pt\hbox{$\sim$}}}\hbox{\raise2pt\hbox{$>$}}}}}

\newcommand\nodata{ ~$\cdots$~ }

\begin{document}

\title[Galaxy Zoo : Bulgeless AGN Host Galaxies]{Galaxy Zoo : Bulgeless Galaxies with Growing Black Holes\thanks{This publication has been made possible by the participation of more than 200,000 volunteers in the Galaxy Zoo project.}} \author[Simmons et al.]{\parbox[t]{16cm}{Brooke D. Simmons$^{1,2,3}\thanks{E-mail: brooke.simmons@astro.ox.ac.uk}$, Chris Lintott$^{1,4}$, Kevin Schawinski$^{2,5}$\thanks{Einstein Fellow}, Edward C. Moran$^{6}$, Anna Han$^{2,5}$, Sugata Kaviraj$^{1,7}$, Karen L. Masters$^{8,9}$, C. Megan Urry$^{2,3,5}$, Kyle W. Willett$^{10}$, Steven P. Bamford$^{11}$, Robert C. Nichol$^{8,9}$
\vspace{0.1in} }\\
$^{1}$Oxford Astrophysics, Denys Wilkinson Building, Keble Road, Oxford OX1 3RH, UK\\
$^{2}$Yale Center for Astronomy and Astrophysics, Yale University, P.O. Box 208121, New Haven, CT 06520, USA\\
$^{3}$Department of Astronomy, Yale University, New Haven, CT 06511, USA\\
$^{4}$Adler Planetarium, 1300 S. Lake Shore Drive, Chicago, IL 60605, USA\\
$^{5}$Department of Physics, Yale University, New Haven, CT 06511, USA\\
$^{6}$Astronomy Department, Wesleyan University, Middletown, CT 06459, USA\\
$^{7}$Blackett Laboratory, Imperial College London, London SW7 2AZ, UK\\
$^{8}$Institute of Cosmology \& Gravitation, University of Portsmouth, Dennis Sciama Building, Portsmouth PO1 3FX, UK\\
$^{9}$SEPnet,\thanks{www.sepnet.ac.uk} South East Physics Network, School of Physics \& Astronomy, University of Southampton, HighÞeld, Southampton SO17 1BJ, UK\\
$^{10}$School of Physics and Astronomy, University of Minnesota, 116 Church St. SE, Minneapolis, MN 55455, USA\\
$^{11}$Centre for Astronomy and Particle Theory, The University of Nottingham, University Park, Nottingham NG7 2RD, UK
   }
  
\maketitle
  
\label{firstpage}
  
\begin{abstract}
The growth of supermassive black holes appears to be driven by {\changed galaxy mergers, violent merger-free processes and/or `secular' processes}. In order to quantify the effects of secular evolution on black hole growth, we {\changed study} a sample of active galactic nuclei (AGN) in galaxies {\changed with a calm formation history free of significant mergers}, a population that heretofore has been difficult to locate. Here we present an initial sample of 13 AGN in massive ($M_\ast \gtrsim 10^{10}~\mmsun $) bulgeless galaxies --- which lack the classical bulges believed inevitably to result from mergers --- selected from the Sloan Digital Sky Survey using visual classifications from Galaxy Zoo. Parametric morphological fitting confirms the host galaxies lack classical bulges; any contributions from pseudobulges are very small (typically $< 5$\%). 
We compute black hole masses for the two broad-line objects in the sample ($4.2 \times 10^6$ and $1.2 \times 10^7~\mmsun $) and place lower limits on black hole masses for the remaining sample (typically $M_{\mathrm{BH}} \gtrsim 10^6~\mmsun $), showing that significant black hole growth must be possible in the absence of mergers {\changed or violent disk instabilities}.

The black hole masses are systematically higher than expected from established bulge-black hole relations. However, if the mean Eddington ratio of the systems with measured black hole masses ($L/L_{\rm{Edd}} \approx 0.065$) is typical, 10 of 13 sources are consistent with the correlation between black hole mass and \emph{total}  stellar mass. That pure disk galaxies and their central black holes may be consistent with a relation derived from elliptical and bulge-dominated galaxies with very different formation histories implies the details of stellar galaxy evolution and dynamics may not be fundamental to the co-evolution of galaxies and black holes.

  \end{abstract}
  

%
%
\section{Introduction}
%
%

Constraining the contribution of mergers to the evolution of the galaxy population is one of the fundamental challenges in modern galaxy formation theory. Galaxies have long been believed to form hierarchically, building up to their observed sizes through a series of mergers \citep{white78,kauffmann93}. The merger history of each galaxy thus contributes significantly to the galaxy's stellar and gas dynamics, and is also thought to drive the co-evolution of a galaxy with its central supermassive black hole \citep[SMBH;][]{sanders88,dimatteo05,croton06,hopkins06b,hopkins08a}. Given such a {\changed fundamental} effect, distinguishing between the effect of mergers and {\changed that of other evolutionary pathways, such as the slow, internal processes collectively known as `secular' evolution}, is difficult. In this paper we present a sample of massive galaxies {\changed chosen to have had no} significant merger in their history, discuss their properties, {\changed and demonstrate for the first time with such a large sample that substantial black hole growth (to $M_{\rm BH} \gtrsim 10^{6-7}~\rm{M}_{\odot}$) is possible without the advent of a significant merger.}

A galaxy's morphology contains signatures of its evolutionary history. In particular, the assembly of massive disk galaxies through mergers inevitably produces a central bulge component \citep[e.g.,][]{toomre77,walker96,hopkins11c,martig12}, {\changed and some merger-free processes such as violent disk instabilities can also form a bulge \citep[e.g.,][]{noguchi99,d_elmegreen04}.} The bulge is dynamically hot, rising vertically above the disk, and has a steeper density profile than an exponential disk \citep{devaucouleurs}. A galaxy lacking a central bulge thus {\changed must have} a formation history free of {\changed violent formation processes. This implies a lack of} significant mergers, with a strong limit on the mass ratio between the main galaxy and any accreting satellite galaxies \citep[$\sim 1:10$;][though \citeauthor{brook12} \citeyear{brook12} suggest the ratio may be {\changed as high as $1:4$}]{walker96,hopkins11c}.

Such bulgeless galaxies, with a purely secular formation history, might be expected to be rare in a hierarchical scenario. The presence amongst the galaxy population of large bulgeless galaxies thus presents a serious challenge to this picture \citep{kormendy10}, as they cannot have undergone a significant merger yet have assembled stellar masses of $M_\ast \gtrsim 10^{10} \rm{M}_{\odot}$. 

Additionally, the well-established {\changed correlations between} galaxies and their central SMBHs {\changed \citep[e.g.,][]{magorrian98,kormendy01,ferrarese00,tremaine02,marconi03,haringrix04} have} led to the prevalence of major-merger-driven theories for black hole-galaxy co-evolution \citep{sanders88,dimatteo05,croton06,hopkins08a}. {\changed However,} a growing body of recent work suggests minor mergers{\changed , cold accretion} and secular processes may be a more typical means of growing a galaxy and its central black hole, both locally \citep[e.g.,][]{greene10b,jiang11b} and at higher redshift \citep[e.g.,][]{simmons11,cisternas11,schawinski11,schawinski12,kocevski12}. 

Owing in part to the compounded rarity of both massive, bulgeless galaxies and active galactic nuclei (AGN), the extent to which a SMBH can grow in the absence of merger processes remains difficult to characterise. {\changed Galaxies lacking classical bulges but hosting AGN have previously been found; these typically have lower stellar masses compared to the general galaxy population, and/or host black holes with relatively low black hole masses (e.g., NGC 4395, \citeauthor{filippenko03} \citeyear{filippenko03}; NGC 3621, \citeauthor{satyapal07} \citeyear{satyapal07}; NGC 4178, \citeauthor{satyapal09} \citeyear{satyapal09},  \citeauthor{secrest12} \citeyear{secrest12}; NGC 3367 and NGC 4536, \citeauthor{mcalpine11} \citeyear{mcalpine11}; NGC 4561, \citeauthor{arayasalvo12} \citeyear{arayasalvo12}). In some cases, these properties are at least in part a direct result of sample selection, as in studies based on samples of low-mass black holes \citep{gh04, gh07a, greene08,greene10b,jiang11a,jiang11b}.}

This paper uses morphological classifications from the Galaxy Zoo\footnote{www.galaxyzoo.org} project \citep{lintott08,lintott11} to construct a sample of bulgeless galaxies that host actively growing black holes. Selecting galaxies that lack classical bulges (as opposed to galaxies with a more varied history of both secular and merger-driven evolution) enables the isolated study of black hole growth in the absence of mergers. {\changed This selection includes optical detection of an AGN but no restriction on its black hole mass.} These galaxies provide a strong challenge to models of galaxy formation, requiring substantial and ongoing secular growth of a central black hole.

We aim to use this rare population to assess whether these galaxies fall on the same galaxy-{\changed black hole} relations seen in galaxies with more merger-driven histories. By comparing upper limits on bulge masses to black hole masses from broad emission lines {\changed and} lower limits on black hole masses using Eddington limits, we assess the sizes to which black holes can grow over their lifetimes due to secular processes alone.
  
Section \ref{sec:finding} describes the methods used to select bulgeless galaxies with growing black holes {\changed from Galaxy Zoo and the Sloan Digital Sky Survey \citep[SDSS;][]{york00}}. Section \ref{sec:sample} presents the sample of host galaxies, with Section \ref{sec:mbh} detailing how the black hole masses and lower limits are calculated. In Section \ref{sec:discussion} we discuss how bulgeless AGN host galaxies inform our understanding of the co-evolution of black holes and galaxies. Throughout this paper, we assume $H_0 = 71~\mathrm{km/s/Mpc}$, $\Omega_M = 0.27$ and $\Omega_\Lambda = 0.73$, consistent with the most recent WMAP cosmology \citep{komatsu11}.

%
%
\section{Finding Bulgeless AGN Host Galaxies in Galaxy Zoo}\label{sec:finding}
%
%

We use visual morphologies drawn from the Galaxy Zoo 2 project, first described in \citet{masters11a}, to assemble a sample of approximately $10,500$ disk galaxies drawn from SDSS which appear bulgeless or nearly bulgeless. We then select a much smaller sample of bulgeless galaxies which host growing supermassive black holes. This section describes first the initial selection of disk galaxies, followed by the AGN identification and hence a conservative selection of bulgeless AGN host galaxies.

Galaxy Zoo volunteers are asked to classify randomly chosen colour images of SDSS systems by clicking buttons in response to a set of descriptive questions arranged into a decision tree. The most relevant here is a question which asks volunteers to classify the bulges of systems already identified as face-on spirals into one of four categories : \textsc{no-bulge}, \textsc{just-noticeable}, \textsc{obvious} and \textsc{dominant}. A full description of the Galaxy Zoo decision tree is given in \citet{masters11a}.

\subsection{AGN Bias: Point Sources Can Mimic Bulges}
\label{sec:bias}

Galaxy Zoo {\changed provides} many independent classifications of each system {\changed because of the large number of citizen scientists participating}. While this approach has many advantages over classification by either an individual or a small group of experts, it is still prone to the biases inherent in morphological classification. These are particularly acute when dealing with small bulges in systems with AGN, where the presence of a nuclear point source can be confused with a central bulge. 
  
In order to investigate the size of this effect on morphological classification, simulated AGN were added to a subsample of images in the most recent iteration of Galaxy Zoo, which uses data drawn from large \textit{Hubble Space Telescope} surveys including GOODS, GEMS and COSMOS. These AGN host galaxy simulations, created using a similar method to those described in Section 3.3 of \citet{simmons08}, are at higher redshifts than the AGN hosts in the SDSS sample discussed here. However, the much higher resolution of the \textit{HST} Advanced Camera for Surveys (ACS) means the simulations cover a similar spatial resolution as SDSS images of galaxies in Galaxy Zoo 2: an ACS image of a galaxy at $z=1$ has the same resolution, in kpc per pixel, as a SDSS image of a galaxy at $z=0.053$. The results of this test of AGN bias on morphological classification are thus directly applicable to the present study{\changed . \citet{koss11}} also find a similar parallel between host galaxy simulations using \emph{HST} images of galaxies at $z \sim 1$ and lower-redshift SDSS images. 
  
The synthetic AGN host morphologies show that the presence of even a faint nuclear point source can significantly affect the visual classification of a galaxy that would otherwise be classified as bulgeless. Among inactive galaxies with a \textsc{no-bulge} classification of at least $80$\%, the addition of a nuclear point source with just $1/50$th the luminosity of the host galaxy decreases the \textsc{no-bulge} classification by at least 50\%; those classifications are instead transferred mainly to the \textsc{just-noticeable} category, such that $\rm{\textsc{no-bulge}} + \rm{\textsc{just-noticeable}} \geq  70\%$ for a bulgeless host galaxy with a faint simulated AGN. When the nuclear point-source luminosity is increased to $1/10$th that of the host galaxy, the \textsc{no-bulge} classification decreases to $10$\% or less, with a corresponding and significant increase in the \textsc{just-noticeable} and \textsc{obvious} bulge classifications. As the AGN luminosity increases with respect to the host galaxy, the visually classified bulge fraction increases substantially. 

While it is well established that parametric morphologies can overestimate a bulge contribution if an AGN is present but not accounted for in a parametric analysis \citep[e.g.,][]{simmons08}, these results from Galaxy Zoo simulations demonstrate that this tendency also appears in visual classifications of AGN host galaxies. Without care being taken to distinguish nuclear activity from a bulge, therefore, a strict selection is likely to reject many truly bulgeless galaxies hosting both unobscured and obscured AGN. We account for this bias using a combination of a more relaxed initial selection, follow-up visual inspection, and finally parametric separation of host galaxy from AGN.

\subsection{Classical Bulges versus Pseudobulges}
\label{sec:pseudobulge}

One particular difficulty in the discovery and analysis of bulgeless galaxies is the distinction between a bulge and a pseudobulge. A classical bulge, as defined by \citet{kormendy04}, is `an elliptical living in a disk', formed by mergers as discussed above, while a pseudobulge still retains signs of having been formed from disk-driven processes. A pseudobulge is, therefore, unlike a bulge in that it is a dynamically cold system, with stars distributed in a disk but with a somewhat steeper density profile than a typical disk. 
Pseudobulges may also be marked by spiral structure or the presence of a nuclear bar or starbursts, and may have profiles similar to the exponentials seen in disks. Their formation mechanism is a matter of debate \citep[e.g.,][]{kormendy10,fisher10} but the presence of a pseudobulge in a galaxy is consistent with a merger-free history. 

The ability to distinguish between pseudobulges and classical bulges is therefore of key importance in determining whether a galaxy is truly bulgeless. Because pseudobulges have light profiles more closely resembling exponential disks than classical bulges, parametric morphological fitting of a pseudobulge with a S\'ersic profile should indicate a more disk-like profile (where an exponential disk has a S\'ersic index of $n = 1$) rather than a classical deVaucouleur bulge (which has $n=4$). {\changed A S\'ersic index of $n=2$ is commonly used as a divider; this criterion has been shown to be reliable for characterising a sample \citep{fisher08}, although there are outliers. While the typical S\'ersic index for a classical bulge is a function of luminosity/mass \citep[e.g.,][]{graham03a,graham08b}, classical bulges with $n \lesssim 2$ are rare for galaxies with masses comparable to this sample. 

We therefore use the S\'ersic index criterion as an initial assessment on the nature of compact host galaxy components in Section \ref{sec:fit}. Further checks on individual objects use the \citet{kormendy77b} relation to assess the location of a compact host galaxy component on the $\langle \mu_e \rangle-r_e$ plane compared to dynamically confirmed classical and pseudobulges (following \citeauthor{gadotti09} \citeyear{gadotti09}; see Section \ref{sec:bulgecheck} below).}

\subsection{Sample Selection}
\label{sec:selection}
  
\subsubsection{Broad Parent Sample}
Galaxies were selected where at least 30 people classified the system as spiral and {\changed not edge-on, and subsequently} answered the question `How prominent is the central bulge, compared with the rest of the galaxy?'  We further require that the combined classifications $\rm{\textsc{no-bulge}} + \rm{\textsc{just-noticeable}} \geq  0.7$, meaning at least 70\% of the weighted classifications fall into either of these categories. 

Additionally, we require that the galaxies are included in the OSSY catalogue \citep*{oh11}, which provides emission and absorption line measurements produced by the GANDALF code \citep{sarzi06}. This produces a parent sample of $10,488$ galaxies with either no bulge or a small bulge. Future work will concentrate on refining this sample to provide an estimate of the population density of truly bulgeless galaxies; this paper concentrate{\changed s} on the subsample which has growing black holes. 

\subsubsection{AGN Selection}

\begin{figure}
\includegraphics[scale=0.42]{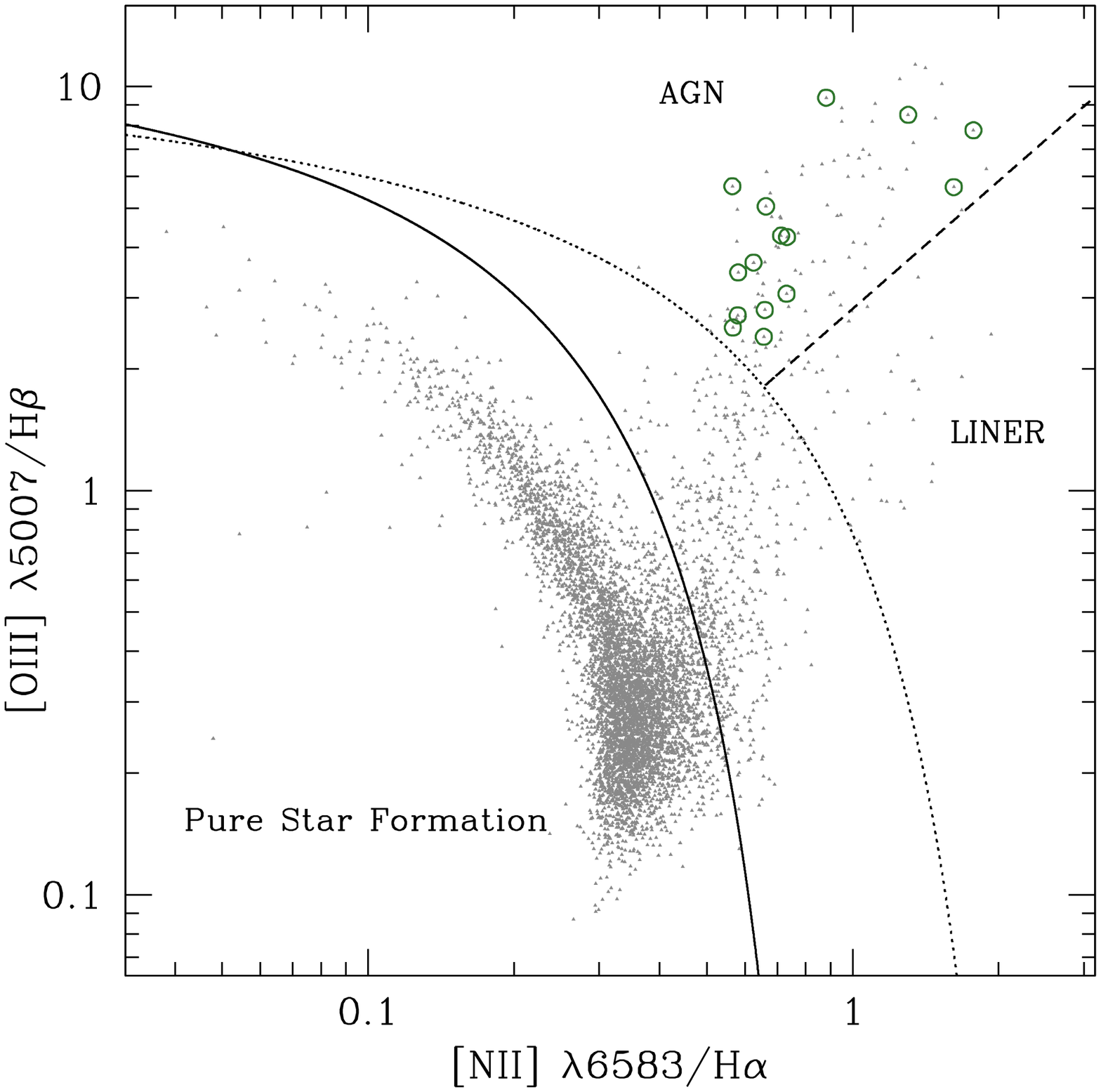}
\caption{
Emission line ratios use{\changed d} as a diagnostic of AGN activity, following \citet{bpt}. The solid line \citep{kauffmann03} empirically separates pure star forming galaxies from composite sources with {\changed both} star formation and AGN activity. The dotted line \citep{kewley01} shows the limit for extreme star formation. The dashed line \citep{schawinski07} shows the empirical AGN-LINER separation. Gray points represent galaxies having a summed Galaxy Zoo 2 classification of $\rm{\textsc{no-bulge}}+\rm{\textsc{just-noticeable}}\geq 70\%$ and emission line $\rm{S/N} \geq 3$ for all four emission lines, from which we selected 15 sources {\changed (open circles)} in the AGN region with no visual evidence of a bulge despite an obvious nuclear point source.
}
\label{fig:bpt}
\end{figure}

Active Galactic Nuclei are selected from the parent sample described above using the optical line diagnostic first described by \citet*{bpt}. Requiring line measurements with signal-to-noise ratio $\rm{S/N} \geq 3$ for each of the \oiii\ $\lambda 5007$, \nii\ $\lambda 6583$, $\mathrm{H\alpha}$, and H$\beta$ lines produces 5,904 sources from the broad parent sample; their positions on the BPT diagram are shown in Figure \ref{fig:bpt}. 

We select galaxies in the region above both the extreme star-formation line of \citet{kewley01} and the empirical AGN-LINER separation of \citet{schawinski07}, which are unambiguously AGN hosts; 100 galaxies lie in this region of the parameter space.

\begin{figure*}
\includegraphics[scale=0.8]{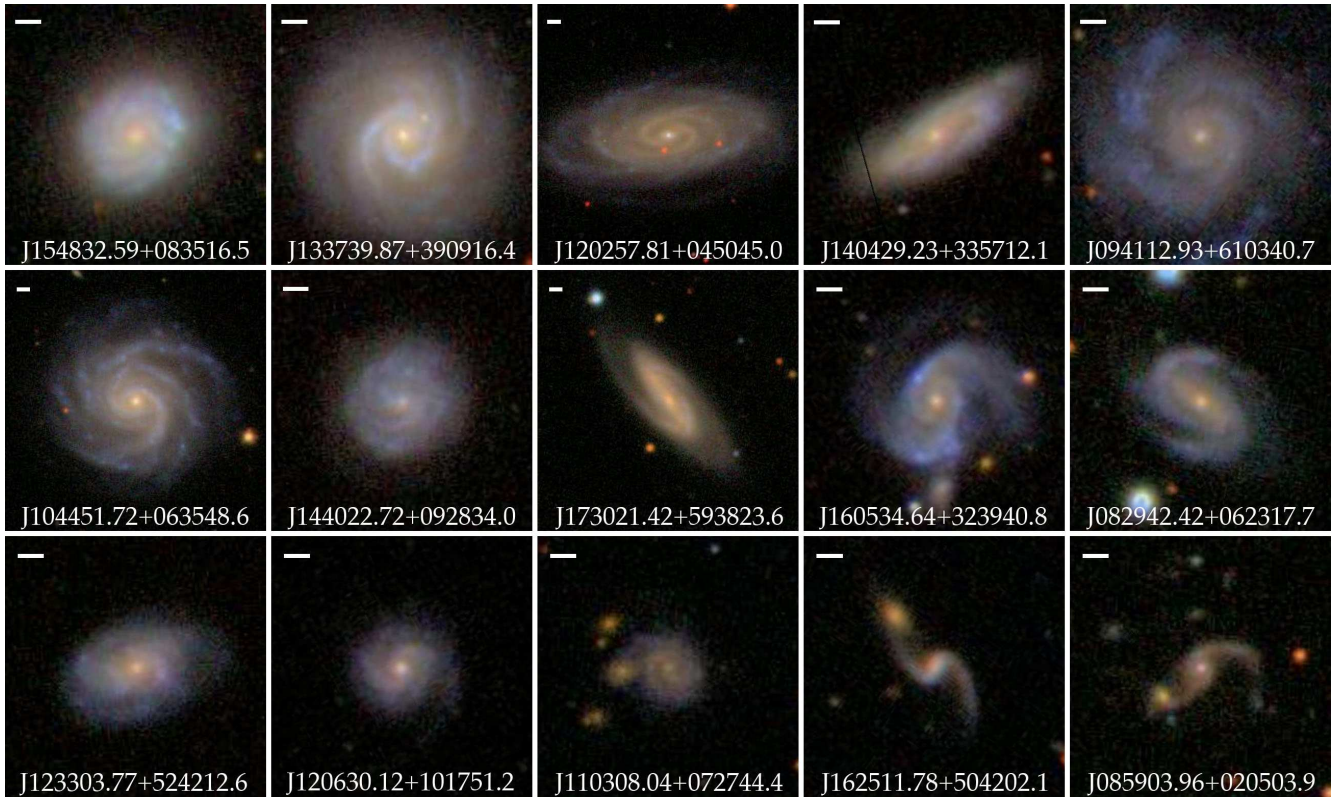}
\caption{
SDSS colour cutouts of 15 AGN (shown in Figure \ref{fig:bpt} as green open circles) with potentially bulgeless host galaxies from visual selection. Reading from top left, the images are sorted by ascending redshift, from $z = 0.014$ to $z=0.19$, matching the order in Table \ref{table:fits}. Each cutout is marked at the top left with a scale bar representing $5''$. {\changed Parametric fitting (described in Section \ref{sec:fit}) reveals the two highest-redshift sources (at the bottom right) to be mergers; the remaining 13 are bulgeless disks.}
}
\label{fig:hostimages}
\end{figure*}

Because this AGN selection method requires strong optical emission from the AGN, all the selected AGN are expected to have visually detectable optical point sources. As the parent sample was selected using broad bulge-classification criteria in order to account for confusion between point-sources and small bulges, many AGN hosts selected will not be truly bulgeless, but will instead have small bulges. To further select a sub-sample of truly bulgeless host galaxies, two authors (BDS and CJL) visually inspected each of the AGN+host galaxy images, selecting only those images with no indication of an extended bulge regardless of the point-like nuclear emission. This very conservative cut produced a sample of 15 AGN host galaxies that appear to be completely lacking a bulge. Figure \ref{fig:hostimages} shows SDSS colour images of each system.

As noted above, this is a lower limit on the AGN fraction as optical selection will not find heavily obscured sources. We have also excluded galaxies in the transition region of the BPT diagram, which include{\changed s} a substantial number of AGN \citep*{trouille11}.

\subsection{Parametric Morphological Fitting}
 \label{sec:fit}
Section \ref{sec:bias} describes the confusion between point sources due to the AGN and any bulge at the centre. In visually inspecting our sample we did not exclude sources with a point-like source at the centre; for a quantitative decomposition between galaxy and point source parametric fitting is necessary \citep{simmons08}. Parametric fitting has the added advantage of enabling both an independent assessment of host morphology and quantitative separation between disk and bulge, providing constraints on possible bulge contribution.

Separation of host galaxy from nuclear emission requires careful characterisation of the image point spread function (PSF). For ground-based observations such as these, the PSF can vary significantly depending on atmospheric conditions, and is difficult to model purely analytically. We therefore use the PSF-creation routines in the IRAF package {\tt daophot} to create a semi-analytical PSF for each image based on stars observed near each system. The number of stars available for each source varies with the source's distance from the galactic plane; between 2 and 40 stars were used for each source (the median number of stars used in PSF creation is 14). Modelling several stars minimizes noise compared to using a single star, whilst still accounting for the unique conditions at each epoch of observation. 

We use the two-dimensional parametric fitting program \galfit\ \citep{peng02,peng10} to simultaneously model the unresolved nucleus and extended galaxy for each of the 15 AGN+host galaxies selected above, choosing the $r$-band images for their depth. {\changed Although $r$ is not immune to dust extinction/reddening, \citet{driver08} and \citet{graham08b} predict the effect of dust on recovered morphological parameters should be minimal in a face-on sample such as ours.} 

Initially, we fit each source's central region with a combination of a single S\'ersic profile \citep{sersic68} and a central point source. Initial parameters (magnitude, radius, axis ratio and position angle) were either drawn from the catalogue or estimated where they are not given by the SDSS catalogue. Initially, the host S\'ersic index is set to $n=2.5$ and allowed to vary. This value was chosen so as to avoid favouring either an exponential disk ($n=1$) or a deVaucouleur bulge ($n=4$). We find, however, that the final best-fit parameters (i.e., with a minimum $\chi^2$) are insensitive to initial indices between $1 \leq n \leq 4$, so long as the other initial parameters are reasonable. 

The primary purpose of this initial fit is to converge on the centroid positions of each component; successive iterations include the extended regions of the galaxy. In order to ensure the extended galaxy is properly fit, we fix the sky background to an independently-determined value for each individual source. Where present, we also fit and subtract nearby bright stars and extended companion galaxies, and mask fainter compact sources from the fit.  

Throughout, the primary goal of the fit was to neither over- nor under-subtract the galaxy's central region. In most cases, this can only be achieved by either masking out asymmetric bright features (such as star-forming knots, bars and spiral arms) or adding them as additional components of the host galaxy fit.  We fit these additional features only when they are necessary to ensure the disk component is properly modelled in the central region of the galaxy.

Once the single S\'ersic plus nuclear point-source fit has converged to its best-fit solution, we add a small second S\'ersic component with a variable $n$ and initially equally bright as the original source, both to constrain the contribution of a small extended bulge that may have been visually obfuscated by the nuclear point source and to distinguish between bulge and pseudobulge. As outlined in Section \ref{sec:pseudobulge}, compact host galaxy components having light profiles with S\'ersic indices $n < 2$ are {\changed typically} considered pseudobulges, whereas components with $n > 2$ are considered classical bulges. 

\begin{figure*}
\includegraphics[scale=0.9]{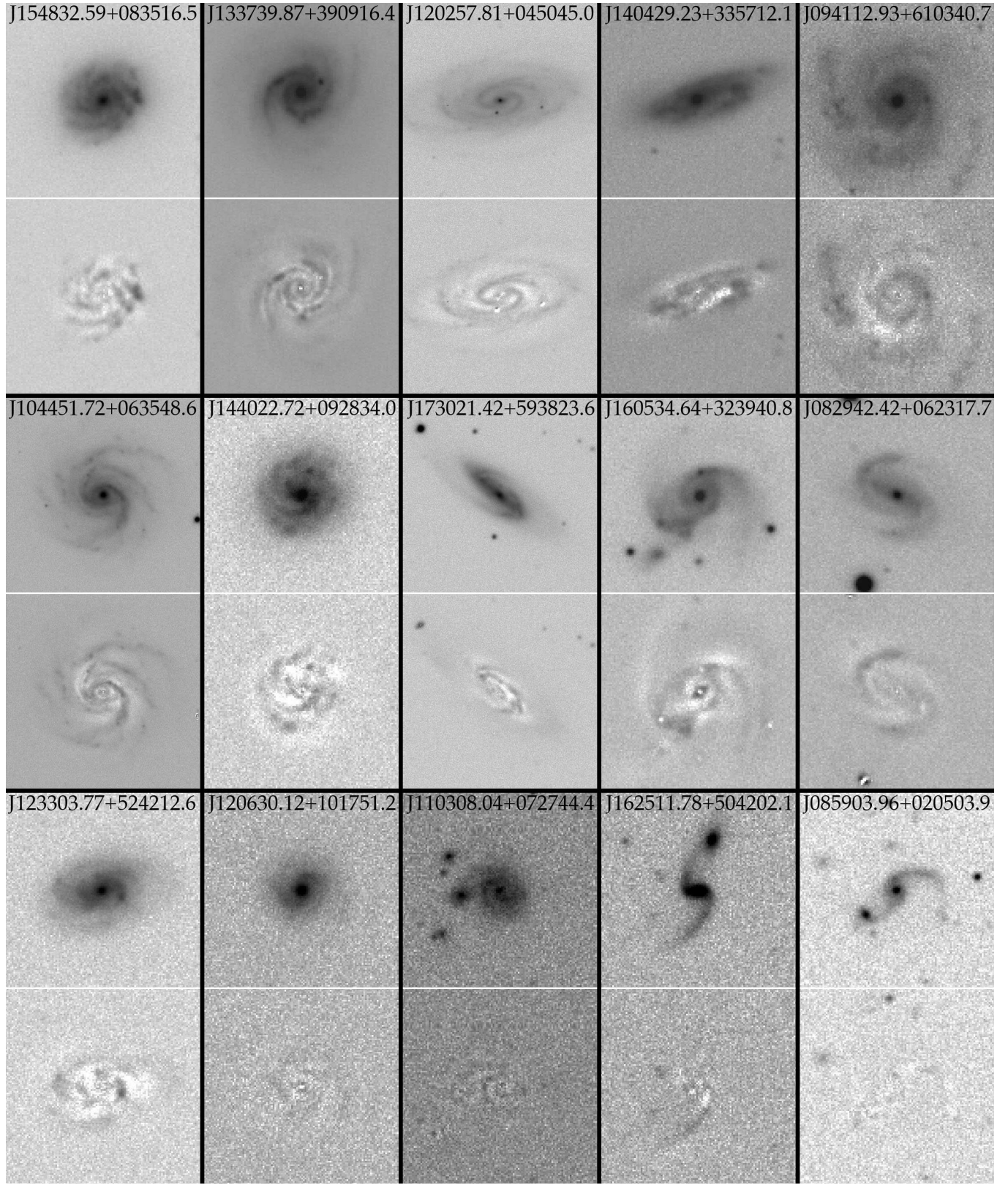}
\caption{
SDSS inverted $r$-band images and residuals from parametric fitting of {\changed the} AGN+hosts in Figure \ref{fig:hostimages}. Images are ordered by ascending redshift (as in Figure \ref{fig:hostimages}); {\changed the} residual after subtracting best-fit model is below each image {\changed (with the same scaling as the original image)}. {\changed Fitting confirms the first 13 galaxies are bulgeless disks;} best-fit parameters are given in Table \ref{table:fits}.
}
\label{fig:hostfits}
\end{figure*}

Figure \ref{fig:hostfits} shows the SDSS $r$-band images and the residual images after subtraction of the best parametric fits. All of the recovered host parameters are reliable because the AGN are significantly fainter than their hosts \citep{simmons08}. In practice, fitting multiple components to a source increases the uncertainties in recovered fit parameters over the computational uncertainties reported by \galfit\ \citep{peng10}. We therefore add additional uncertainties to those reported by \galfit\ using results from extensive parametric host galaxy fitting simulations \citep{simmons08}. This additional uncertainty particularly affects the faint, compact second host galaxy components, but where a compact host component is detected we can nevertheless distinguish between bulge and pseudobulge in all but one system (described in Section \ref{sec:sample}). However, when calculating the upper limit to a possible bulge contribution to the galaxy (such as in Figure \ref{fig:mbh_mstar}), we include all the light from even those compact host components firmly detected as pseudobulges. We therefore consider our bulge limits conservative upper limits. 


\begin{table*}
\begin{tabular}{|c|c|c|c|c|c|c|c|c|c|}
\hline
SDSS ID
 & Redshift & \multicolumn{2}{c}{Extended Host S\'ersic(s)} & \multicolumn{2}{c}{Compact Host S\'ersic} & \multirow{2}{*}{$\frac{L_\mathrm{{compact}}}{L_\mathrm{{host, tot}}}$} & \multirow{2}{*}{$\frac{L_\mathrm{{PS}}}{L_\mathrm{{host, tot}}}$} & AGN $L_\mathrm{{bol}}$ & $\log M_\mathrm{{BH}}$\\
&  & $n$ & $r_e$ [pix] & $n$ & $r_e$ [pix] &  & & [erg s$^{-1}$] & [$\mmsun $]\\
\hline
J154832.59+083516.5 & $0.0144$ & $ 0.92^{+ 0.05}_{- 0.05} $ & $ 21.30^{+ 0.44}_{- 0.44} $ & $ 1.17^{+ 0.24}_{- 0.24} $ & $ 3.49^{+ 0.34}_{- 0.34} $ & $ 0.029 $ & $ 0.003 $ & $ 43.7 $ & $> 5.6 $ \\ 
& & $0.21^{+ 0.05}_{- 0.05}$ & $17.48^{+ 0.12}_{- 0.12}$ & & & & & & \\
J133739.87+390916.4 & $0.0198$ & $ 0.82^{+ 0.03}_{- 0.03} $ & $ 26.31^{+ 0.04}_{- 0.04} $ & $ 0.53^{+ 0.20}_{- 0.20} $ & $ 3.50^{+ 1.00}_{- 1.00} $ & $ 0.011 $ & $ 0.014 $ & $ 44.1 $ & $ 7.1 \pm 0.13$ \\ 
J120257.81+045045.0 & $0.0207$ & $ 0.67^{+ 0.03}_{- 0.03} $ & $ 65.07^{+ 0.22}_{- 0.22} $ & $ 0.52^{+ 0.06}_{- 0.06} $ & $ 4.81^{+ 0.12}_{- 0.12} $ & $ 0.022 $ & $ 0.012 $ & $ 43.4 $ & $ 6.6 \pm 0.14 $ \\ 
J140429.23+335712.1 & $0.0264$ & $ 0.43^{+ 0.20}_{- 0.20} $ & $ 26.54^{+ 0.04}_{- 0.04} $ & $ 1.57^{+ 0.22}_{- 0.22} $ & $ 4.00^{+ 0.90}_{- 0.90} $ & $ 0.075 $ & $ 0.007 $ & $ 43.9 $ & $> 5.8 $ \\ 
J094112.93+610340.7 & $0.0265$ & $ 0.70^{+ 0.03}_{- 0.03} $ & $ 38.55^{+ 0.14}_{- 0.14} $ & $ 2.04^{+ 0.72}_{- 0.72} $ & $ 4.50^{+ 1.00}_{- 1.00} $ & $ 0.037 $ & $ 0.006 $ & $ 43.4 $ & $> 5.3 $ \\ 
J104451.72+063548.6 & $0.0276$ & $ 0.89^{+ 0.02}_{- 0.02} $ & $ 41.00^{+ 0.08}_{- 0.08} $ & $ 0.70^{+ 0.08}_{- 0.08} $ & $ 2.66^{+ 0.06}_{- 0.06} $ & $ 0.038 $ & $ 0.008 $ & $ 44.4 $ & $> 6.3 $ \\ 
J144022.72+092834.0 & $0.0282$ & $ 0.53^{+ 0.03}_{- 0.03} $ & $ 24.91^{+ 0.15}_{- 0.15} $ & $ 1.47^{+ 0.20}_{- 0.20} $ & $ 4.63^{+ 0.46}_{- 0.46} $ & $ 0.046 $ & $ 0.008 $ & $ 43.6 $ & $> 5.5 $ \\ 
J173021.42+593823.6 & $0.0284$ & $ 0.62^{+ 0.03}_{- 0.03} $ & $ 44.04^{+ 0.25}_{- 0.25} $ & $ 0.24^{+ 0.32}_{- 0.19} $ & $ 4.40^{+ 0.64}_{- 0.64} $ & $ 0.026 $ & $ 0.006 $ & $ 44.3 $ & $> 6.2 $ \\ 
J160534.64+323940.8 & $0.0297$ & $ 1.00^{+ 0.03}_{- 0.03} $ & $ 25.32^{+ 0.18}_{- 0.18} $ & $ 1.39^{+ 0.48}_{- 0.48} $ & $ 3.00^{+ 0.90}_{- 0.90} $ & $ 0.055 $ & $ 0.022 $ & $ 43.7 $ & $> 5.6 $ \\ 
J082942.42+062317.7 & $0.0516$ & $ 0.73^{+ 0.03}_{- 0.03} $ & $ 38.51^{+ 0.58}_{- 0.58} $ & $ 0.11^{+ 0.40}_{- 0.06} $ & $ 2.81^{+ 0.50}_{- 0.50} $ & $ 0.037 $ & $ 0.029 $ & $ 44.1 $ & $> 6.0 $ \\ 
J123303.77+524212.6 & $0.0557$ & $ 0.81^{+ 0.03}_{- 0.03} $ & $ 24.85^{+ 0.73}_{- 0.73} $ & $ 0.13^{+ 1.46}_{- 0.08} $ & $ 1.67^{+ 1.52}_{- 0.67} $ & $ 0.025 $ & $ 0.016 $ & $ 44.4 $ & $> 6.3 $ \\ 
J120630.12+101751.2 & $0.0635$ & $ 1.13^{+ 0.06}_{- 0.06} $ & $ 19.10^{+ 0.21}_{- 0.21} $ & \nodata & \nodata & $< 0.02 $ & $ 0.069 $ & $ 44.0 $ & $> 5.9 $ \\ 
J110308.04+072744.4 & $0.0850$ & $ 0.47^{+ 0.03}_{- 0.03} $ & $ 12.96^{+ 0.12}_{- 0.12} $ & \nodata & \nodata & $< 0.02 $ & $ 0.017 $ & $ 43.4 $ & $> 5.3 $ \\ \hline
J162511.78+504202.1 & $0.1279$ & $ 2.65^{+ 0.18}_{- 0.18} $ & $ 15.09^{+ 0.46}_{- 0.46} $ & \nodata & \nodata & $< 0.02 $ & $ 0.059 $ & $ 44.8 $ & $> 6.7 $ \\ 
J085903.96+020503.9 & $0.1889$ & $ 0.60^{+ 0.26}_{- 0.26} $ & $ 11.99^{+ 0.73}_{- 0.73} $ & \nodata & \nodata & $< 0.02 $ & $ 0.156 $ & $ 45.2 $ & $ 8.2 $ \\
\hline
\end{tabular}
\caption{
Properties of the 15 sources initially selected from Galaxy Zoo as potentially bulgeless AGN+host galaxies. After parametric morphological fitting, we retain the first 13 sources in the sample. All sources have an extended host galaxy component and unresolved nuclear point source. 11 of 13 sources also have a resolved, but compact, nuclear host galaxy component. Ten compact host components have $n < 2$ and the remaining has $n \sim 2${\changed ; all 11 are more than $3\sigma$ below the $\langle \mu_e \rangle-r_e$ relation of \citet{kormendy77b} for classical bulges \citep{gadotti09} and are thus classified as pseudobulges. The mean and median contributions of the pseudobulges to their host galaxy light are 3.6 and 3.3\%, respectively}. The first source in the table, J154832.59+083516.5, has two extended disks \emph{and} a small pseudobulge. Most of the black hole masses are lower limits calculated by assuming the black hole is radiating at the Eddington limit; the three sources in the table with broad $\mathrm{H\alpha}$ emission lines have firm black hole masses, but J085903.96+020503.9 is {\changed removed} from the sample due to a merging companion.}\label{table:fits}
\end{table*}

By construction, the sample is unambiguously disk-dominated. However, one source (J162511.78+504202.1) appears to be a merger of a galaxy with a strong bulge ($n=2.65 \pm 0.18$) and a companion, with tidal tails that resemble spiral arms. Another (J085903.96+020503.9) is visually similar, but fitting indicates a disk-dominated ($n = 0.6 \pm 0.26$) central component with a bulge-dominated companion. This system contains a broad-line AGN, and fitting the extended arms requires strong Fourier (asymmetric) modes. As the photometric redshift of the companion is consistent with a physical interaction between it and the primary source, this is likely a merger or post-merger and the arms may in fact be tidally induced features. Both galaxies are removed from our sample. 

%
%
\section{Sample properties}\label{sec:sample}
%
%

\subsection{Are these AGN host galaxies really bulgeless?}
\label{sec:bulgecheck}

The remaining sample of 13 host galaxies are all well fit by a model consisting of a dominant disk and a nuclear point source, providing strong constraints on the maximum contribution of a small bulge component. In 3 cases, we do not detect a second, compact host component. For the other systems, examination of residuals from fitting only the extended disk + nuclear point-source components shows clear signs of a small extended component in the center; in all but one case this additional component has a Sersic index consistent with a pseudobulge ($n < 2$ within the $1 \sigma$ uncertainties). The mean and median contributions of these pseudobulges to the total host galaxy light are 3.6\% and 3.3\%, respectively.

The sole exception (J094112.93+610340.7) has a marginal $n = 2.0 \pm 0.7$, meaning we cannot say {\changed from this criterion alone} whether it is a classical bulge or a pseudobulge. {\changed However, its mean surface brightness within the effective radius, $\langle \mu_e \rangle$, is lower than the $3 \sigma$ lower bound for the $\langle \mu_e \rangle-r_e$ relation for classical bulges and elliptical galaxies \citep[Figure 8 of][]{gadotti09}, strongly suggesting it is indeed a pseudobulge. (The other compact host components in the sample also lie in the pseudobulge region below the classical bulge region in the $\langle \mu_e \rangle-r_e$ plane.)} We also note that the host galaxies with no detected pseudobulge are the highest-redshift sources in the sample; higher-resolution imaging could clarify both the status of this one exception and the pseudo/bulgeless nature of those sources with $z > 0.06$. 

{\changed As an additional check on the robustness of the fits, we examined near-infrared images for those objects currently covered in the UKIDSS \citep{lawrence07} $K$-band. Only four galaxies currently have sufficient depth for reliable separation of host galaxy components using the techniques described in Section \ref{sec:fit}, and all 4 have morphological parameters consistent with those obtained for the $r$-band images. This supports the assumption in Section \ref{sec:discussion} that the mass-to-light ratios of the disks and pseudobulges are not significantly different and indicates that dust extinction is not preferentially causing a loss of pseudobulge flux in this sample.

It is therefore highly likely that these galaxies are truly bulgeless. However, we conservatively assume that all of the light from each pseudobulge component could be light from a classical bulge and consider it a robust upper limit on the contribution of a classical bulge.} When no bulge is detected, we assume the upper limit to be 2\% of the host galaxy light. The minimum detected pseudobulge contribution is 1\%; the maximum is 7.5\%.

\subsection{Host Galaxy Properties}

\begin{figure}
\includegraphics[scale=0.68]{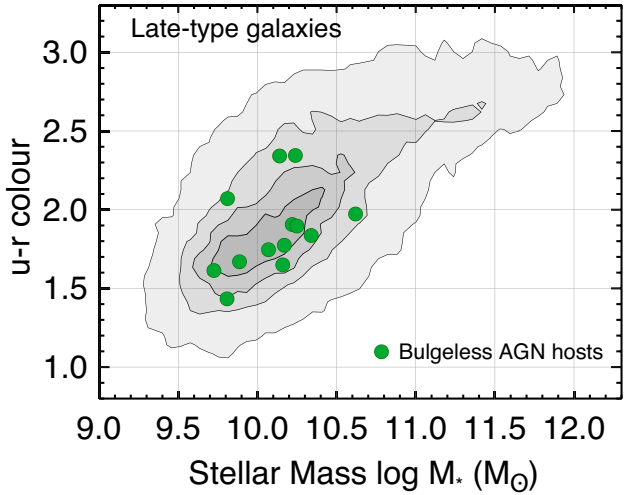}
\caption{
$u-r$ colour versus stellar mass for SDSS inactive late-type {\changed galaxies} with $0.02 < z < 0.05$ \citep[black and grey contours;][]{schawinski10a}, with the 13 bulgeless AGN host galaxies from this work shown in green. The bulgeless host galaxies span a range of masses and colours but are mainly located in the blue cloud at masses typical of the inactive disk galaxies from \citeauthor{schawinski10a} However, as the AGN host sample presented here is incomplete and was not selected in the same way as the inactive late-type sample, we caution against detailed comparisons between these {\changed samples}.
}
\label{fig:cmd}
\end{figure}

The sample is insufficiently large to draw significant conclusions about host galaxy properties, but it is worth noting that the galaxies, as seen in Figure \ref{fig:hostimages}, are heterogenous examples of disk galaxies, comprising both barred and unbarred, red and blue, and tightly and loosely wound spirals. Even with this small sample, it is obvious that bulgeless galaxies hosting AGN are not restricted to a single morphological class. They span typical stellar mass and colour values for inactive late-type galaxies (Figure \ref{fig:cmd}). This sample does not contain host galaxies firmly on the red sequence; {\changed however, a detailed comparison} to the inactive population is difficult owing to our emphasis on a pure rather than a complete sample. Disk galaxies on the red sequence are more likely to host growing black holes than those in the blue cloud \citep{masters10c}; further work is required to determine the fraction of bulgeless disk galaxies (both hosting AGN and not) on the red sequence, as well as the overall population of bulgeless galaxies.

Table  \ref{table:fits} summarizes the results of parametric fitting. The $L_{\rm{AGN}}/L_{\rm{host}}$ ratios range from 0.003 to 0.069. Extended S\'ersic profiles are disk-like (median $n = 0.73$), and the disk effective radii range from 2.4 to 15.2 kpc.
 
One galaxy (J154832.59+083516.5) is best fit with two extended disks, which have similar axis ratios ($b/a \approx 0.8$) but are rotated by approximately $16^\circ$ with respect to one another. The two disks are visually evident in the colour image (top left panel of Figure \ref{fig:hostimages}), and the fit improves with the addition of the second extended disk component: the reduced goodness-of-fit parameter is $\chi^2_\nu = 1.032$ with two disks versus 1.244 with a single disk (with $\sim 40,000$ degrees of freedom), a significant improvement in the fit.

We note that the faint \emph{apparent} companions to two of the galaxies shown in Figure \ref{fig:hostimages} (J160534.64+323940.8 and J110308.04+072744.4) are in fact projections of more distant background galaxies, according to photometric redshifts. J160534.64+323940.8 also has an asymmetric spiral arm pattern, but asymmetric spiral features are not necessarily signatures of interaction, as disk instabilities can lead to bars, warps and other asymmetric features without external interaction \citep{saha07,sellwood10}.

\citet{baldry06} derive a useful environmental measure for SDSS galaxies closer than a redshift of 0.085. The local density for a galaxy is given by $\Sigma_{N}=N/\left(\pi d_N^2\right)$ where $d_N$ is the projected distance to the Nth nearest neighbour that is more luminous than $M_r=20$, and $\Sigma$ is determined by averaging the density determined using spectroscopic neighbours with that from using both photometric and spectroscopic neighbours. We use the extension to SDSS DR6 described in \citet{bamford09}. 12 of our sample  are included in the catalogue; values range from $\Sigma=0.076$ to 2.48 with a mean of 0.450, corresponding to the density of a typical field environment. For comparison, the entire SDSS has typical local density measures from 0.2 to 25. The environment{\changed s} of our systems {\changed are} therefore markedly different from {\changed those} of merging galaxies \citep{darg10b}, supporting the idea that these bulgeless systems are free from recent merger activity. 

Stellar masses were derived following \citet{baldry06} using a best fit stellar mass-to-light ratio corrected for the observed dependence on $u-r$ colour. This approach, which was first introduced by \citet{bell01}, is not as accurate as one based on full spectral fitting, but retains a simple relation between observed and derived quantities. It is also less likely to be distorted by the presence or absence of an AGN with luminosities like those in this sample.

%
%
\section{Black Hole Masses}\label{sec:mbh}
%
%

\begin{figure}
\includegraphics[scale=0.45]{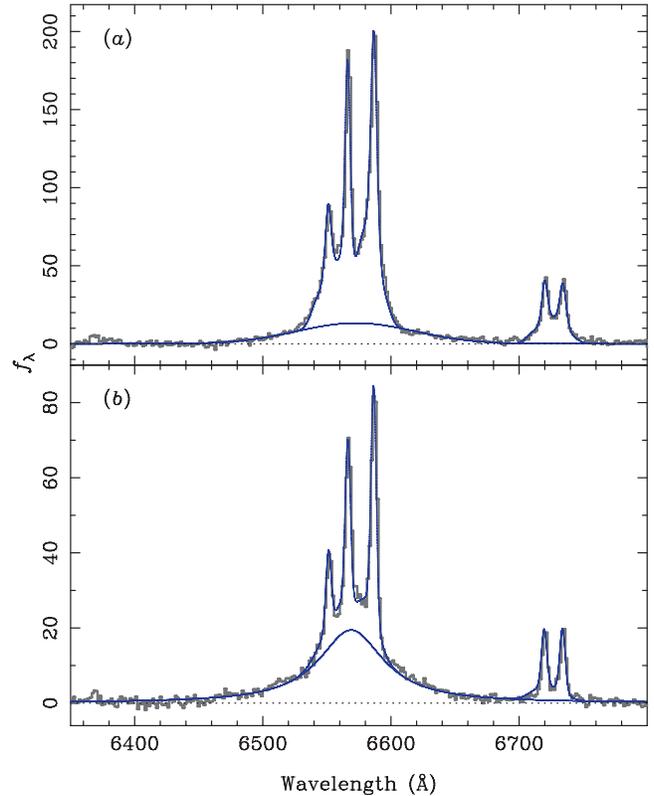}
\caption{
SDSS spectra (dark gray) of (a) J120257.81+045045.0 and (b) J133739.87+390916.4, with spectral fits to narrow and broad components (blue). Narrow \nii , \sii , and $\mathrm{H\alpha}$ are fit separately in order to measure the isolated flux and width of the broad $\mathrm{H\alpha}$ component. 
}
\label{fig:blagn}
\end{figure}

Two of the AGN in the sample (J120257.81+045045.0 and J133739.87+390916.4) have broad emission lines. We used the $\mathrm{H\alpha}$ line width and flux to measure the black hole masses of these sources, following \citet{gh07a} and \citet{jiang11b}. 
We began by fitting the stellar continuum present in each spectrum using the {\tt GANDALF} software \citep{sarzi06}.  Although {\tt GANDALF} is designed to fit both absorption and emission lines, the profiles of the narrow emission lines in both objects are complex and asymmetric, so we instead subtracted the continuum fits from the data and analyzed the residual emission-line spectra ourselves.  The process involved (a) modeling the profile of the \oiii\ 5007 line as the sum of 2--3 Gaussian components, (b) fitting the \oiii\ model to the \sii\ 6716,31, \nii\ 6548,83, and narrow H$\alpha$ lines, and (c) modeling the residual broad H$\alpha$ components as a single Gaussian (J133739.87+390916.4) or Lorentzian (J120257.81+045045.0) profile.  The results are shown in Figure \ref{fig:blagn}.  For J133739.87+390916.4, the broad H$\alpha$ line has a width of 4950~km/s FWHM and a luminosity of $1.38\times 10^{40}$~erg/s, which correspond to a black-hole mass of $1.2 \times 10^7~\mmsun$ \citep{jiang11b}.  The broad H$\alpha$ line in J120257.81+045045.0 has a width and luminosity of 2810~km/s and $1.9 \times 10^{40}$~erg/s, respectively, suggesting a black-hole mass of $4.2 \times 10^6 ~\mmsun$.  A Gaussian fit to the broad H$\alpha$ component of this object would imply a slightly larger black-hole mass, although it is inappropriate based on the spectral fits (Figure \ref{fig:blagn}).

Lower limits on black hole masses for the remaining sample can be obtained {\changed from} the bolometric luminosity of the sources. For sources like these where the host galaxy dominates the emission at most wavelengths, bolometric luminosities are typically obtained using corrections to either the hard X-ray or mid-infrared bands, where the AGN emission dominates. All the sources are detected by the Wide-field Infrared Survey Explorer \citep[\emph{WISE};][]{wright10} with $\rm{S/N} > 5$. We use the wavelength-dependent bolometric corrections of \citet{richards06}, which are not strongly dependent on AGN luminosity, to estimate the bolometric luminosity based on the longest-wavelength infrared data available for each source. Twelve of 13 AGN with bulgeless hosts are detected in the W4 band centered at $22~\mu$m ($L_{\rm{bol}} \approx 10 \times L_{22 \mu\rm{m}}$); all are detected in the W3 band centered at $12~\mu$m ($L_{\rm{bol}} \approx 8 \times L_{12 \mu\rm{m}}$). We verified that black hole mass limits calculated in this way are not significantly different from mass limits for this sample calculated using a bolometric correction to the \oiii\ luminosity {\changed \citep{heckman04, stern12}}.

Given the bolometric luminosity, making the assumption that accretion is at the Eddington limit yields a \emph{lower} limit on the black hole mass. Super-Eddington accretion is required for a black hole mass limit derived in this way to be overestimated; this is only rarely observed and, when present, typically results in only a small change in observed luminosity \citep{collin04}. We note that the two sources with measured black hole masses are accreting at rates well below their Eddington limits ($L_{\rm bol}/L_{\rm Edd} =  0.05$ and 0.08 for J120257.81+045045.0 and J133739.87+390916.4, respectively). If these growth rates are typical of the sample, it implies the actual black hole masses of the sample are higher by at least $\sim 1$~dex than the computed lower limits. The lower limits obtained on black hole mass are substantial, ranging from $\sim 10^{5}$ to $10^{6}~\mmsun$. The black hole masses/mass limits are given in Table \ref{table:fits}.

%
%
\section{Discussion}\label{sec:discussion}
%
%

\begin{figure*}
\includegraphics[scale=0.8]{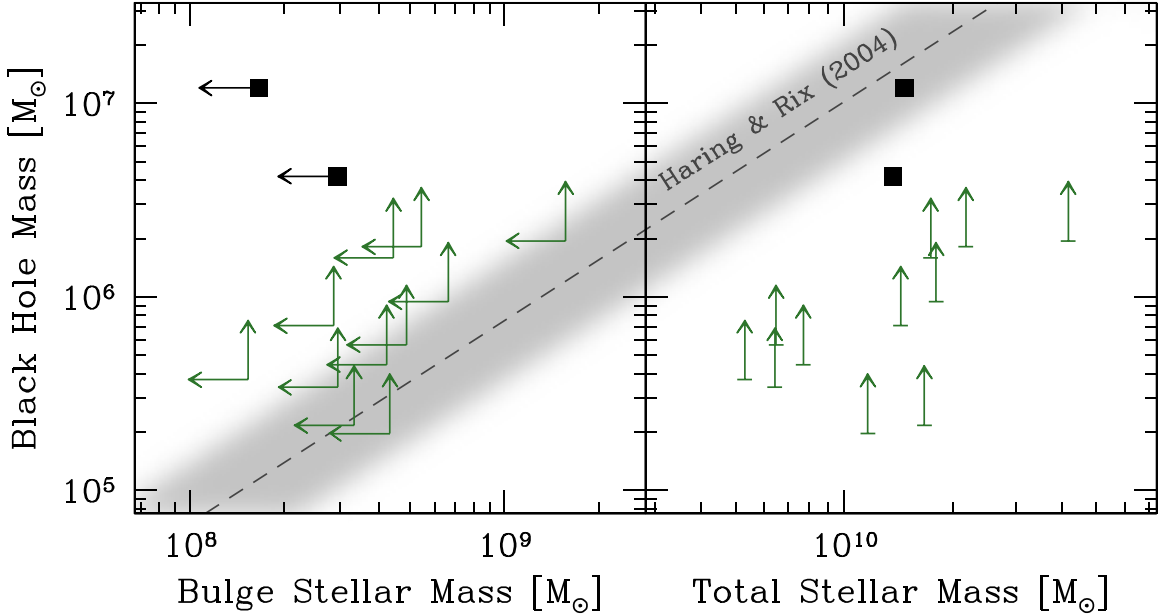}
\caption{
Black hole mass versus bulge mass (left) and total host galaxy stellar mass (right). Host galaxy stellar masses are calculated following \citet{baldry06}. {\changed No source has a detected classical bulge; we calculate robust upper limits to bulge masses (left panel; leftward arrows) using the $r$-band pseudobulge fraction}. AGN with broad-line emission have black hole masses calculated from $\mathrm{H\alpha}$ line width and flux (black squares). Vertical arrows indicate lower black hole mass limits calculated assuming radiation at the Eddington limit; super-Eddington accretion is required for a black hole to have a mass below the limit. The local relation between bulge/elliptical stellar mass and black hole mass from \citet{haringrix04} is plotted as a dashed line; the shaded region indicates the observed scatter of 0.3 dex quoted by \citeauthor{haringrix04}. The black hole masses are higher than predicted by that relation based on the maximum bulge masses of these systems. However, all black hole masses/limits are consistent with the relation \emph{if} the relation describes total stellar mass. If the Eddington ratios for the two systems with measured black hole masses are typical of the sample, the limits underestimate the black hole masses by at least $\sim 1$~dex, amplifying the offset in the left panel, but bringing the sample into good agreement with the correlation between black hole mass and \emph{total} stellar mass {\changed on the right}.
}
\label{fig:mbh_mstar}
\end{figure*}

Massive pure disk galaxies hosting growing supermassive black holes offer a powerful probe of the co-evolution of galaxies and black holes in the absence of {\changed bulge-building mechanisms. However, relatively few such systems are currently known. The challenges involved in finding them arise} from a combination of factors. {\changed Firstly, galaxy evolution as a whole is currently thought to have proceeded \citep[e.g.,][]{white78,springel05a,maccio10} via processes favouring the formation of significant bulges, including mergers and violent disk instabilities such as clumps \citep[][though also see \citeauthor{inoue11} \citeyear{inoue11}]{noguchi99,d_elmegreen04,b_elmegreen08}. In such a universe, massive bulgeless disks formed in the absence of these processes} are expected to be very rare \citep[e.g.,][]{steinmetz02}. Additionally, only a small fraction of those are expected to host actively growing supermassive black holes at any given time. Identification of AGN via optical emission lines is possible over volumes large enough to enclose a rare host galaxy sample, but such diagnostics miss optically obscured AGN; the nuclear emission from those AGN that are detected can lead to visual confusion with a small bulge, impeding the identification of bulgeless galaxies hosting optically-selected AGN. 

However, using a combination of Galaxy Zoo visual classifications within the Sloan Digital Sky Survey and parametric morphological analysis, we have selected 13 such galaxies. The unambiguously disk-dominated sample presented here was selected to favour purity rather than completeness and can thus be considered a robust initial sample of massive 
AGN host galaxies without classical bulges and with very small (or undetected) pseudobulges.


The contribution of pseudobulges to these systems is extremely small even compared to other samples of AGN host galaxies with histories thought to be dominated by secular evolution \citep{orbandexivry11,mathur12} or to samples of inactive massive bulgeless galaxies \citep{kormendy10}. Of the bulgeless giant galaxies discussed by \citeauthor{kormendy10}, only the extremely disk-dominated (3 of 19 in that sample) have comparable pseudobulge-to-disk ratios. {\changed The galaxies in our sample are thus fully consistent with being pure-disk AGN host galaxies, which places strong constraints on their histories: they have grown to masses of $M_\ast \sim 10^{10}~\mmsun$ in the absence of any significant mechanism building classical \emph{or} pseudobulges. } 

In particular, these galaxies have no major mergers in their collective history, and their minor merger history is strongly limited. Simulations typically show that mergers with mass ratios greater than $1:10$ form a classical bulge \citep{walker96,hopkins11c}; the lack of bulges in this sample constrains its merger history to events with a smaller ratio (in the sense that the satellite is less than $1/10$th the main galaxy's mass). 

Note that some recent work suggests bulges may be suppressed in mergers up to a mass ratio of $\sim 1:4$ \citep{brook12}, which, if true, would allow a galaxy to remain bulgeless after merging with somewhat larger galaxies than the typical limit from most simulations. However, such a significant minor merger should leave signs of tidal debris; such signs could be detectable in images at the depths shown here, depending on the stage of relaxation \citep[e.g., the more minor mergers in the spheroidal post-merger sample of][]{carpineti12}. None of the galaxies in the sample show evidence of tidal features at the SDSS depth, disfavouring the notion that a minor merger event precipitated the observed black hole growth {\changed \citep[deeper imaging could confirm this;][]{mihos05,rudick09,kaviraj10,martinezdelgado10,cooper10}}. 

Additionally, \citeauthor{brook12} predict long-lived bar features in bulgeless galaxies that have undergone a minor merger, but no excess of bars is observed in the sample, suggesting this is not a significant effect. {\changed This is consistent with recent results showing no clear link between bars and black hole growth \citep{oh12,lee12}, even though galactic-scale bars are linked to the overall evolution of disk galaxies \citep{masters11a, skibba12}. As bars are also linked to the growth of pseudobulges \citep[e.g.,][]{bureau99,kormendy04,athanassoula05,fisher09}, the extremely low pseudobulge-to-total ratios strongly suggest these galaxies have had a very calm evolutionary history for the majority of a Hubble time, with neither major mergers nor episodic gas accretion that could be considered violent even at $z > 2$ \citep{cooper10,martig12}. Neither major nor minor mergers, nor even merger-free processes that create sizable pseudobulges,} are driving the observed black hole growth.

Given the bolometric luminosities of the sample, and assuming the bolometric luminosity $L_{\mathrm{bol}}$ (energy radiated) is related to the accretion rate $\dot m$ (energy captured) by a radiative efficiency factor $\epsilon$, one can estimate the amount of matter falling onto each black hole. Adopting a value of $\epsilon = 0.15$ \citep{elvis02} yields accretion rates of $0.003 \lesssim \dot{m} \lesssim 0.03~\mmsun\mathrm{~yr^{-1}}$ for the sample. This level is easily achievable with feeding via dynamically cold accretion of minor satellites \citep{crockett11}, but it may also be possible with merger-free processes alone. For example, \citet{ciotti91} calculated that stellar mass loss from a passively evolving stellar population (albeit in an elliptical galaxy) could send material toward a central SMBH (see \citeauthor{ciotti12} \citeyear{ciotti12} for a recent treatment, and \citeauthor{davies07} \citeyear{davies07} for examples in local Seyfert galaxies). The amount of material that could be driven toward the nucleus within a disk galaxy remains uncertain and the details of such a process are unclear, but gas-rich disks with higher star formation rates and the increased potential for transfer of angular momentum present many purely secular opportunities for feeding central supermassive black holes \citep[for a detailed review, see][]{jogee06}. {\changed Given the very small pseudobulges in our sample, however, secular processes that grow both pseudobulges \emph{and} black holes \citep[such as violent disk instabilities;][]{schawinski11b,bournaud11,bournaud12} must still be limited.}

Using black hole accretion rate and mass, one can estimate the time required to grow a seed SMBH to the observed mass. We assume the black holes have grown at the rate (i.e., the luminosity) currently observed since the mass at which that rate was the Eddington rate, and assume Eddington-limited accretion for mass growth that occurred prior to that point. This effectively means the observed accretion rate is assumed to be the maximum rate (in $\mmsun\mathrm{~yr^{-1}}$) at which the black hole has grown over its lifetime. {\changed This is quite conservative given the low rates observed}, but it provides an estimate of the total time a black hole of the given mass would need to spend in an actively growing phase over its lifetime if the observed rate is typical. For the two systems with firm black hole masses, the time required is between $\sim 1- 2 \mathrm{~Gyr}$, depending on the seed mass \citep[the range given assumes seed masses between $10^2$ and $10^5 ~\mmsun$;][]{volonteri08, volonteri10}. 

Thus even if the currently observed accretion rate is the maximum rate over the lifetime of these AGN, the time required to grow SMBHs to the masses observed here is considerably less than a Hubble time. The SMBHs in bulgeless galaxies need only spend $\sim 10$\% of their lifetimes in an actively growing phase, a similar fraction to that predicted by independent models and observed by others \citep[e.g.,][]{kauffmann03,hao05,hopkins06c,fiore12}. This does not rule out the possibility of higher accretion rates in the past (e.g., with cold accretion flows at high redshift) and less time spent in an active growth phase, but such a phase is not necessary. Further characterisation of {\changed past black hole growth} requires firm black hole masses for the remainder of the sample. 

Whatever process has fed these SMBHs, it is clear that the formation of these very small pseudobulges does not correlate their properties with $M_{BH}$ in the same way as that of classical bulges. Figure \ref{fig:mbh_mstar} compares black hole mass with the contribution to the stellar mass of the host galaxy from a bulge or pseudobulge component (under the assumption that the mass-to-light ratio of the compact galaxy component is equal to that of the whole galaxy), for comparison to measured galaxy-black hole correlations. Such correlations \citep[e.g.][]{marconi03,haringrix04} compare black hole masses to classical bulge properties. {\changed However, there are no detected classical bulges in the sample. We therefore calculate maximum bulge masses by using the mass of the pseudobulge as a strong upper limit on the mass of a classical bulge.} Comparison with the measured correlation between these quantities {\changed (Figure 6)} shows that the black hole masses allowed by the derived limits are systematically above the measured correlation with bulge stellar mass. 

{\changed This result differs from several previous studies \citep[e.g.,][]{greene08,jiang11a,mathur12} finding that black holes in galaxies with pseudobulges are smaller than predicted by standard bulge-black hole relations. However, this apparent conflict is a result of different sample selection techniques. Whereas our study starts by selecting bulgeless host galaxies and then examines black hole masses, previous studies of pseudobulges and/or bulgeless host galaxy samples have typically started by selecting low-mass SMBHs and subsequently examined the host galaxies. In studying narrow-line Seyfert 1 galaxies, for example, \citet{mathur12} preferentially selected lower-mass central black holes undergoing Eddington-limited accretion. Similarly, the sample of \citet{gh07a} was created by first selecting unobscured AGN with detected broad optical emission lines and then limiting the black hole mass to $M_{BH} < 2 \times 10^6~\mmsun$. The visual selection method used in this paper, on the other hand, may be biased \emph{against} selecting broad-line AGN unless a more aggressive correction for bulge-nucleus confusion is used. As a result, there is no overlap between this sample and that of \citeauthor{gh07a}, despite the two being drawn from the same optical dataset. 

It may not be surprising that the host galaxies in different samples are more similar than the black holes, especially between studies using varying selection methods within the same parent dataset. Taken as an ensemble, the collective results may be evidence that there is no intrinsic correlation between pseudobulge mass and black hole mass \citep[in agreement with][]{kormendy11a}, but a more uniform selection of a larger sample may be necessary to confirm that. We note, for example, that the typical galaxy in this sample has a lower redshift than the sub-sample of \citet[][which uses the sample of \citeauthor{gh07a} \citeyear{gh07a}]{jiang11b} with B/Tot~$< 5\%$, as well as a slightly higher total stellar mass.
}

Figure \ref{fig:mbh_mstar} also shows the relationship between black hole mass and total stellar mass, for comparison to the same black hole-galaxy mass relation. Note that the relation is based on elliptical/bulge masses for primarily bulge-dominated systems (26 of 30 galaxies in \citeauthor{haringrix04} \citeyear{haringrix04} are classified as E or S0), so $M_{\rm bulge} \approx M_{\rm *,tot}$ for most of the systems on which the relation of \citeauthor{haringrix04} is based, in contrast to the total stellar masses of the bulgeless galaxies in this sample. Because the disks in this sample have very different dynamical histories than elliptical or bulge-dominated galaxies, we do not expect their masses to correlate with black hole mass in the same way as bulges if different dynamical histories lead to different rates of black hole growth. 

However, the comparison of bulgeless galaxies to the correlation based on bulge-dominated galaxies shows very good agreement. One of the two systems in the sample for which absolute measures of black hole mass are available is consistent with the relation, while the other has a black hole mass just below the observed scatter (0.3 dex; but given the uncertainties in the stellar masses, it is marginally consistent). The systems with limits on the black hole masses are consistent with the relation, but given the Eddington ratios of the two systems with measured black hole masses, these limits are likely underestimates by at least 1 dex. If we assume the Eddington ratios of the two systems with measured masses are typical of the sample and apply them to the black hole mass estimates of the remaining 11 systems, 9 of the 11 fall within the scatter of the \citet{haringrix04} relation between black hole mass and \emph{total} galaxy stellar mass (the remaining two are outside the scatter by approximately the same amount as J120257.81+045045.0). That the black hole masses of pure disk galaxies may correlate with total stellar host galaxy mass in the same way as SMBHs in bulge-dominated and elliptical galaxies -- despite very different galactic formation histories -- indicates that the evolutionary processes driving the dynamical and morphological configuration of the galaxy stellar mass may not be fundamental to the growth of the central black hole. 

The results presented here should be read in the context of results from a growing number of simulations which show that the black hole-galaxy connection is a reflection of mutual correlations between these two components and the overall gravitational potential of the dark matter halo \citep{booth10,volonteri11}, such that the black hole-galaxy relation is a natural outcome of hierarchical galaxy evolution \citep{jahnke11} regardless of the merger history of the galaxy. The observational evidence is less clear: some work indicates a correlation between halo mass and black hole mass \citep{bandara09}, with some evidence that outliers to the $M-\sigma$ relation are not outliers on a similar relation between SMBH and dark matter halo \citep{bogdan12}, but others find no correlation between black hole mass and dark matter halo \citep{kormendy11b}.

Regardless of whether the galactic-scale evolution of baryons is fundamental to the evolution of supermassive black holes, bulgeless galaxies hosting AGN provide a means of studying merger-free and/or dynamically cold pathways to supermassive black hole growth in relative isolation compared to the majority of galaxies with a more complicated history of mergers and secular processes. They may eventually provide leverage to constrain the maximum black hole growth possible via merger-free processes in all galaxies, and to separate the extent to which SMBH and galaxy properties correlate as a result of different evolutionary processes. The results presented here indicate that significant black hole growth (to at least $\sim 10^7~\mmsun$) is possible via pathways free of mergers, although a more complete treatment requires follow-up work to determine firm black hole masses for the remainder of the sample.

%

%
%
\section{Conclusions}
%
%

By using classifications of visual morphologies of Sloan Digital Sky Survey galaxies, drawn from the Galaxy Zoo project, we have selected a large set of bulgeless face-on spiral galaxies. A conservative initial selection identifies 13 of these galaxies which are unambiguously systems with growing black holes. Parameterized fitting of these galaxies provides stringent limits on bulge or pseudobulge mass, with the latter typically contributing $\sim3$\% by mass on average. 

Two of the galaxies in the sample have broad-line AGN, and thus measurements of their black hole mass are possible. For the rest, infrared observations from the \emph{WISE} mission allow us to place a lower limit on the black hole mass. The black hole masses are substantial, reaching $\sim10^7~\mmsun$, and lie above those predicted by the local bulge-black hole mass relation, even when all the pseudobulge component is included as an upper limit to the mass of the classical bulge in each case.

One of the two black holes with measured masses is fully consistent with the relation between black hole and total stellar mass (the other is just outside the scatter). If the Eddington limits of the black holes with measured masses are typical of the full sample, $80\%$ of the systems for which only lower limits are available have black hole masses consistent with predictions based on \emph{total} galaxy stellar mass. Firm conclusions require further observations, but it is not inconsistent with the idea that black hole mass is more closely related to the overall gravitational potential of the galaxy and its dark matter halo (which is dominated by the halo but traced by the total stellar mass) than to the dynamically hot bulge component.
  
In any case, the presence of massive, growing super-massive black holes in bulgeless galaxies indicates that secular evolution is an important part of the evolution of the galaxy population. Either significant black hole growth is possible even in the absence of significant {\changed bulge-building mechanisms}, or a dynamical means to keep galaxies bulgeless despite these mechanisms must be found. Future work will include an analysis of the more than 10,000 candidate bulgeless galaxies from which this sample was drawn in order to constrain the properties of this intriguing population; in particular, a search for bulgeless systems in mergers will distinguish between the two scenarios left open. Observational follow-up of the small sample identified here, particularly in order to constrain more tightly the bulge properties and black hole masses is also urgently necessary. Although extending the work to higher redshift will be challenging, Galaxy Zoo classifications for large \emph{Hubble Space Telescope} studies hold the promise of identifying a similar set of galaxies out to a redshift of approximately one. Even at low redshift, however, the systems identified here present a stringent test of simulations of galaxy formation.

%
%
\section*{Acknowledgments}
%
%

The authors wish to thank to C. Peng for making \galfit\ publicly available, and for many enlightening discussions.  We also wish to thank R. Skibba, M. Williams, and the anonymous referee for thorough and constructive comments that helped us improve this manuscript.
The JavaScript Cosmology Calculator \citep{wright06} and TOPCAT \citep{taylor05} were used while preparing this paper. 
BDS acknowledges support from Worcester College, Oxford, and from NASA through grant HST-AR-12638.01-A.
Support for the work of KS was provided by NASA through Einstein Postdoctoral Fellowship grant number PF9-00069, issued by the Chandra X-ray Observatory Center, which is operated by the Smithsonian Astrophysical Observatory for and on behalf of NASA under contract NAS8-03060.
ECM acknowledges support from the National Science Foundation through grant AST-0909063.
SK acknowledges fellowships from the 1851 Royal Commission, Imperial College London, Worcester College, Oxford and support from the BIPAC institute at Oxford.
KLM acknowledges funding from The Leverhulme Trust as a 2010 Early Career Fellow.
SPB acknowledges receipt of an STFC Advanced Fellowship.
RCN acknowledges STFC Rolling Grant ST/I001204/1 to ICG for ÔSurvey Cosmology and AstrophysicsÕ. 

Galaxy Zoo was supported by The Leverhulme Trust. 

This publication makes use of data products from the Wide-field Infrared Survey Explorer, which is a joint project of the University of California, Los Angeles, and the Jet Propulsion Laboratory/California Institute of Technology, funded by the National Aeronautics and Space Administration.

Funding for the SDSS and SDSS-II has been provided by the Alfred P. Sloan Foundation, the Participating Institutions, the National Science Foundation, the U.S. Department of Energy, the National Aeronautics and Space Administration, the Japanese Monbukagakusho, the Max Planck Society, and the Higher Education Funding Council for England. The SDSS Web Site is http://www.sdss.org/.

The SDSS is managed by the Astrophysical Research Consortium for the Participating Institutions. The Participating Institutions are the American Museum of Natural History, Astrophysical Institute Potsdam, University of Basel, University of Cambridge, Case Western Reserve University, University of Chicago, Drexel University, Fermilab, the Institute for Advanced Study, the Japan Participation Group, Johns Hopkins University, the Joint Institute for Nuclear Astrophysics, the Kavli Institute for Particle Astrophysics and Cosmology, the Korean Scientist Group, the Chinese Academy of Sciences (LAMOST), Los Alamos National Laboratory, the Max-Planck-Institute for Astronomy (MPIA), the Max-Planck-Institute for Astrophysics (MPA), New Mexico State University, Ohio State University, University of Pittsburgh, University of Portsmouth, Princeton University, the United States Naval Observatory, Yale University and the University of Washington. 
  
\bibliographystyle{mn2e}
\bibliography{refs}

\end{document}